\input harvmac.tex
%\input epsf.tex
%\draft
%\newcount\figno
\figno=0
\def\fig#1#2#3{
\par\begingroup\parindent=0pt\leftskip=1cm\rightskip=1cm\parindent=0pt
\baselineskip=11pt
\global\advance\figno by 1
\midinsert
\epsfxsize=#3
\centerline{\epsfbox{#2}}
\vskip 12pt
{\bf Fig. \the\figno:} #1\par
\endinsert\endgroup\par
}
\def\figlabel#1{\xdef#1{\the\figno}}
\def\encadremath#1{\vbox{\hrule\hbox{\vrule\kern8pt\vbox{\kern8pt
\hbox{$\displaystyle #1$}\kern8pt}
\kern8pt\vrule}\hrule}}

\overfullrule=0pt

%macros
%

\def\np#1#2#3{{\it Nucl. Phys.} {\bf B#1} (#2) #3}
\def\pl#1#2#3{{\it Phys. Lett. }{\bf B#1} (#2) #3}
\def\prl#1#2#3{{\it Phys. Rev. Lett. }{\bf #1} (#2) #3}
\def\physrev#1#2#3{{\it Phys. Rev.} {\bf D#1} (#2) #3}

\font\zfont = cmss10 %scaled \magstep1

\def\bigone{\hbox{1\kern -.23em {\rm l}}}
\def\ZZ{\hbox{\zfont Z\kern-.4emZ}}

\def\a{\alpha}
\def\b{\beta}

\def\k{\kappa}

\def\m{\mu}
\def\n{\nu}

\def\D{\Delta}

\def\o{\over}

\Title{CALT-68-2195, hep-th/9810050}
{\vbox{
\hbox{\centerline{Quantum Gravity Corrections for}} 
\hbox{\centerline{Schwarzschild Black Holes }}
}}
\smallskip
\centerline{Katrin Becker\footnote{$^\diamondsuit$}
{\tt beckerk@theory.caltech.edu} and Melanie 
Becker\footnote{$^\star$}
{\tt mbecker@theory.caltech.edu} }
\smallskip
\centerline{\it California Institute of Technology 452-48, 
Pasadena, CA 91125}
\bigskip
\baselineskip 18pt
\noindent

We consider the Matrix theory proposal describing eleven-dimensional Schwarzschild black holes. 
We argue that the Newtonian potential between two black holes
receives a genuine long range quantum gravity correction, which is finite 
and can be computed from the supergravity point of view.
The result agrees with Matrix theory up to a numerical factor
which we have not computed.

\Date{October, 1998}

\newsec{Introduction}
Matrix theory
\ref\bsss{T.~Banks, W.~Fischler, S.~H.~Shenker and 
L.~Susskind, ``M Theory as a Matrix Model: A Conjecture'', 
\physrev {55} {1997} {5112}, hep-th/9610043.} 
is the non-perturbative formulation of
the eleven dimensional M-theory.
The non-perturbative nature of black hole physics indicates
that Matrix theory might be an ideal framework to address questions
in black hole physics.
A concrete model describing Schwarzschild black holes 
in diverse
dimensions was proposed by
 Horowitz and Martinec
\ref\hm
{G.~T.~Horowitz and E.~J.~Martinec, ``Comments on Black Holes in
Matrix theory'', 
\physrev {57} {1998} {4935}, hep-th/9710217.} and 
Banks, Fischler, Klebanov and Susskind 
\ref\bfks{T.~Banks, W.~Fischler, I.~R.~Klebanov and L.~Susskind,
``Schwarzschild Black Holes in Matrix Theory.II'', 
{\it J.~High Energy Phys.} {\bf 01} (1998) 008, hep-th/9711005.}
which describes the black hole in terms of a Bolzmann gas of 
distinguishable D0-branes.
This model reproduces correctly the Bekenstein-Hawking entropy\foot{A different proposal for a microscopic derivation 
of the Bekenstein-Hawking entropy for Schwarzschild black 
holes is presented in 
\ref\ss{K.~Sfetsos and K.~Skenderis, ``Microscopic Derivation of the 
Bekenstein-Hawking Entropy Formula for Non-Extremal
Black Holes'', \np {517} {1998} {179}, hep-th/9711138. }.},
the size, the mass, as well as the leading term of the static Newtonian
 potential 
between two black holes.
Liu and Tseytlin 
\ref\lt{H.~Liu and A.~Tseytlin, ``Statistical Mechanics
of D0-Branes and Black Hole Thermodynamics'', 
{\it J.~High Energy Phys.} {\bf 01} (1998) 010, hep-th/9712063.} 
have generalized this picture 
to include all loop large N supersymmetric Yang-Mills (SYM) corrections, or
equivalently corrections from general relativity.

In this paper we shall be interested in the Newton potential
between a pair of Schwarzschild black holes described by the model 
{\hm}, {\bfks} and {\lt}. Our main interest is the existence of a genuine
quantum gravity correction to the gravitational potential, which is 
{\it finite} and
can be computed from the supergravity point of view. 
It is the leading long range quantum gravity correction
that can be computed from Matrix theory as well.
Happily, we will see that both
results indeed agree up to a numerical factor that we have not computed.

The existence of finite quantum gravity corrections to the 
four-dimensional Newtonian
potential has been emphasized by
Donoghue
\ref\do{J.~F.~Donoghue,
''General Relativity as an Effective Field Theory:
The Leading Quantum Corrections'', \physrev {50} {1994}
{3874}, gr-qc/9405057, ``Leading Quantum Correction to the 
Newtonian Potential'', \prl {72} {1994} {2996}, gr-qc/9310024.}.
In a series
of interesting papers he  
showed that the leading long distance 
quantum gravity correction to Newton's potential can be reliably calculated 
in a quantum theory of gravity.
Let us briefly review the general idea.

The leading term of 
the four-dimensional Newtonian potential for
two particles with mass $M_1$ and $M_2$ separated by a 
distance $r$ is given by
\eqn\ai{V(r)=-G {M_1 M_2 \o r},}
where $G$ is Newton's constant. 
This expression is of course only approximately valid.
For large masses or large velocities there are relativistic corrections
which can be computed in a post-Newtonian expansion and 
have been verified experimentally. 
For a small test particle $M_2$ this can be seen from the expansion of 
the time component of the Schwarzschild metric:
\eqn\aii{g_{00}={{1-{GM_1 \o r c^2} \o 1+ {GM_1 \o r c^2}}\approx
1-{2 GM_1 \o r c^2} \left( 1- { G M_1 \o r c^2}+\dots\right). }}

In a quantum theory of gravity we expect that the potential
will be corrected by quantum effects.
It is well known that when trying to unify 
quantum mechanics and general relativity we will
face the problem that this is a non-renormalizable theory.
Although one can quantize the theory on smooth enough 
backgrounds the divergences appearing in particular diagrams 
are such that they cannot be absorbed into the coupling constants
of a minimal general relativity. 
If one introduces new coupling constants 
to absorb the divergences one is led to an infinite number of free
parameters and thus to a lack of predictivity. 
Despite this situation the leading
long distance quantum corrections can be reliably
calculated in `quantum' general relativity.

To leading order in the distance
two massive objects will interact through a
Newtonian potential of the form:
\eqn\aiii{V(r)=-{G M_1M_2 \o r} \left( 1+ a{G(M_1+M_2) \o rc^2}+ b{G \hbar \o r^2c^3}+\dots\right),}
where $a$ and $b$ are some finite numerical coefficients.
The first correction is due to general relativity which  
is the term appearing in \aii. 
The correction proportional to $\hbar$ is a true quantum gravity effect. 
Its overall form can be fixed by dimensional analysis while
the numerical coefficient 
can be calculated if
we treat gravity as an effective field theory where we
can make an expansion in the energy.
This procedure is familiar from chiral perturbation
theory which represents the low energy limit of QCD.

The action describing gravity is then organized in terms
of powers of the curvature

\eqn\dii{S_{grav}=\int d^4 x \sqrt{-g}\left(
{2\o {\k}^2}R+\a R^2+\dots\right),}
where $\k^2 \sim G$ and $\a$ is a constant.
Derivatives are related to the momentum $\partial _{\m} \sim p_{\m}$, 
so that the curvature is of order $p^2$, two powers of curvature are of
order $p^4$, and so on. At low energies or equivalently long
distances terms of order $p^4$ can be ignored compared terms of
order $p^2$.

The quantum gravity correction {\aiii} is the leading long 
distance quantum gravity effect which is due to the long range 
propagation of 
massless particles such as the graviton.
Only the lowest order coupling contained in the Einstein
action is needed to describe this effect 
since higher order interactions like $R^2+\dots$
contain too many powers of the momentum and are neglible at large distances.
Since this quantum correction to {\aiii} is finite it 
might be the first and simplest quantum gravity corrections that 
could actually be tested by Matrix theory. This is the purpose
of our paper.

There are certainly many contributions 
to the effective action in a quantum theory of gravity that 
are divergent like for example the $R^4$ contribution considered in
\ref\ggv{M.~B.~Green, M.~Gutperle and P.~Vanhove, 
``One Loop in Eleven dimensions'', \pl {409} {1997} {177}, hep-th/9706175.}
\ref\ep{E.~Keski-Vakkuri and P.~Kraus, ``Short Distance Contributions
to Graviton-Graviton Scattering: Matrix Theory versus Supergravity'',
\np {529} {1998} {246}, hep-th/9712013.}
\ref\bbr{K.~Becker and M.~Becker, ``On Graviton Scattering 
Amplitudes in M-Theory'', \physrev {57} {1998} {6464}, hep-th/9712238.}.
These are not the effects that we would like to consider here. 
Matrix theory regularizes the
divergent contributions in M-theory by fixing the value of the cutoff
and there is no finite result on the gravity side 
with which one could compare.

Let us now come to the Matrix theory version of the story.
A number of important papers suggested that Schwarzschild black
holes could be described in string theory and Matrix theory
\ref\bhp{L.~Susskind, ``Some Speculations about Black Hole
Entropy in String theory'', hep-th/9309145, 
G.~T.~Horowitz and J.~Polchinski, ``A Correspondence Principle 
for Black Holes and Strings'', \physrev {55} {1997} {6189}, 
hep-th/9612146, T.~Banks, W.~Fischler, I.~R.~Klebanov and 
L.~Susskind, ``Schwarzschild Black Holes from Matrix Theory'', 
\prl {80} {1998} {226}, hep-th/9709091, 
I.~R.~Klebanov and L.~Susskind, ``Schwarzschild Black Holes
in Various Dimensions from Matrix Theory'', 
\pl {416} {1998} {62}, hep-th/9709108.}.
Horowitz and Martinec {\hm} and
Banks, Fischler, Klebanov and Susskind {\bfks} have proposed a beautiful 
model
whose aim is to describe Schwarzschild black holes in 
diverse dimensions. Many interesting features of black hole
physics are correctly reproduced by this model.
The basic idea is that Schwarzschild black holes in diverse dimensions
can be described in terms of a system of D0-branes in toroidally
compactified space.
We shall be interested in the eleven-dimensional case. One might have 
thought that this case is especially difficult, because very little is known
about the SYM theory for general $N$.
However, as we shall see, enough is known to determine the
quantum gravity correction we are interested in.

In this paper we would like to see if the Matrix theory description
is able to reproduce the leading 
long distance
quantum gravity correction to Newton's potential of an 
eleven-dimensional Schwarzschild black hole.
We will use crude methods in the sense that we will not compute 
any numerical coefficients. 
All the dependences turn out to work out correctly so that
precise agreement  
between Matrix theory and supergravity can be found.  

This paper is organized as follows. 
In the section 2 we would like to consider the Matrix model point of view. 
We will recapitulate
the result for the leading term of Newton's potential and
compute relativistic corrections.
We will see that the leading long distance quantum gravity 
effect follows from a one loop term in the Matrix model. 
In section 3 we consider the supergravity side of the story, 
which we then compare with Matrix theory. 
Finally, in section 4 we will present our 
conclusions.

\newsec{The Newtonian Potential from Matrix Theory}
Let us start by reviewing some relevant features of the Matrix theory
model for Schwarzschild black holes.
From now on we will only take into account general
dependences but not any numerical coefficients.
The basic idea of {\hm} and {\bfks}
is that a $D$-dimensional Schwarzschild black hole 
can be described
in terms of a 
Boltzmann gas of distinguishable D0-branes with two body 
interactions given by the 
leading term in the one-loop 
SYM effective action compactified on $T^d$ $(D=11-d)$. We will be 
interested in the case $D=11$ where the Lagrangian is

\eqn\bi{
L_{eff}={N v^2 \o R}+{ G_{11} N^2 \o R^3}{v^4 \o r^{7}},}
where $G_{11}$ is the eleven-dimensional Newton's constant, 
$R$ is the radius of the eleven dimension, $N$ is the number of D0-branes
and $v$ and $r$ are the relative
velocity and distance respectively.
This Lagrangian has a holographic scaling property {\bsss}. We would like to 
rescale
the relative transverse coordinate $r$ and the time coordinate $t$
as
\eqn\bii{
\eqalign{
& r \rightarrow N^{1 \o 9} r,\cr 
& t \rightarrow {1 \o R} N^{2 \o 9} t. \cr }}
From {\bii} it follows that the velocity gets rescaled as
\eqn\biii{v\rightarrow R (NG_{11})^{-{1 \o 9}}v.}
The dependence on Newton's  constant in formula {\biii}
can be determined by dimensional analysis.
The Lagrangian {\bi} becomes (up to an overall factor) 
independent of $N$ and $R$.
Applying a qualitative analysis based on the virial theorem, 
i.e. equating the kinetic and potential term of {\bi}
one can derive a relation 
between $N$ and the size of the bound state 

\eqn\biv{ G_{11}^{-1} R_s^{9}=N.}
Since one is treating the D0-branes as a Boltzmann
gas its entropy is of order $N$ so that the above
relation is precisely the Bekenstein-Hawking area law for a 
Schwarzschild black hole {\hm}, {\bfks} and 
\ref\li{
M.~Li, ``Matrix Schwarzschild Black Holes in the Large $N$ Limit'',
{\it J.~High Energy Phys.} {\bf 01} (1998) 009, hep-th/9710226.},
where $R_s$ corresponds to the Schwarzschild radius.
Using the standard relation between the 
light cone energy and mass one can determine the 
scaling of the mass $M$ as a function of $N$

\eqn\biv{M \sim G_{11}^{-1/9}N^{8/9}.}
This means the Matrix model correctly reproduces the 
scaling of the mass in terms of the Schwarzschild radius.

Finally, one can show that the static Newtonian 
potential between two equal mass black holes can be
be completely understood in terms
of the velocity dependent potential between the D0-branes.
For this purpose we use the connection between the light cone energy
and the rest mass plus the potential {\bi} and scaling relations 
that we just mentioned.
The result for the (leading term of the) eleven-dimensional Newton
potential is
{\bfks}
\eqn\biv{
V(r)\sim G_{11} {M^2 \o r^{7}}(G_{11}N)^{-{1 \o {9}}}=
G_{10} {M^2 \o r^{7}}.}
The last factor on the left hand side of {\biv}
comes from the averaging over the longitudinal 
direction. This is because this factor is $1/R_s$ and if one 
is working for $S \sim N$ then one can set $R_s \sim R$. 
The reason for this is that the longitudinal box 
expands in such a way that the black hole fits into it
(see the discussion in {\bfks} and {\hm}).
We have further used  $G_{10} \sim G_{11}/R$.

Let us now discuss general relativity corrections to the previous
gravitational potential.
Liu and Tseytlin 
{\lt}
have proposed a generalization of the
 action {\bi} which includes all loop large $N$ SYM corrections. 
The basic idea is to consider the classical Born-Infeld action
for a D0-brane probe moving in a supergravity background
produced by a D0-brane source.
The all loop large $N$ SYM corrections to {\bi} can be obtained if 
one formally extrapolates this action to the short distance or 
near horizon region.
The Lagrangian obtained in this way is
\eqn\bv{L_{eff}={r^7R \o G_{11}} \left[ \sqrt{1-G_{11}{N \o R^2} {v^2 \o r^7}} -1 \right].}
Expanding {\bv} one gets {\bi} plus corrections
\eqn\bvi{L_{eff}={N v^2 \o R} +{G_{11} N^2 \o R^3} {v^4 \o r^7}+
{G_{11}^2 N^3 \o R^5} {v^6 \o r^{14}} +\dots.}

Let us consider the third term appearing in {\bvi}.
If we follow the same steps
as for the leading contribution we conclude
that this expression corrects the gravitational potential 
in the following way
\eqn\bvii{\D V(r) \sim G^2_{10} {M^3 \o r^{14}}.}
This is precisely the form of a relativistic correction to
Newton's potential in $D=11$ as can be seen for example by analogy
to the four dimensional formula {\aiii} that we mentioned 
in the introduction\foot{Of course $c=1$ in string 
theory conventions.}.
By the same argument it is easy to see
that higher order corrections in {\bvi}
reproduce the correct form of higher order relativistic
corrections to Newton's potential.

Let us now come to our main point: the leading long 
distance quantum gravity correction.
By now it is well known that the effective action of Matrix theory
at a given number of loops is a double expansion in 
the relative velocity and distance between D0-branes.
The leading one loop term, i.e. the $v^4$-term appearing in {\bi}, 
is not renormalized at higher loops 
\ref\bb1{K.~Becker and M.~Becker, ``A Two-Loop Test of M(atrix) 
Theory'', \np {506} {1997} {48}, hep-th/9705091.}
\ref\sethi1{S.~Paban, S.~Sethi and M.~Stern, 
``Constraints from Extended Supersymmetry in Quantum
Mechanics'', hep-th/9805018.}.
The $v^6$-term has a very special property: it vanishes at one
loop and it is not renormalized beyond two loops 
\ref\sethi2{S.~Paban, S.~Sethi and M.~Stern, ``Supersymmetry and Higher 
Derivative Terms in the Effective Action of Yang-Mills Theories'',
{ \it J.~High Energy Phys.} {\bf 06} (1998) 012, hep-th/9806028.}.
From these considerations we come to the conclusion that the 
leading long distance quantum gravity correction to
the one-loop $v^4$-term 
corresponds to the $v^8$-term at one loop. 
Let us have a closer look at this contribution. 

From the systematic expansion derived in
\ref\bbpt{K.~Becker, M.~Becker, J.~Polchinski and A.~Tseytlin,
''Higher Order Graviton Scattering in M(atrix) Theory'', \physrev {56} {1997}
{3174}, hep-th/9706072.} the concrete form 
of this term is
\eqn\bviii{\D L_{eff}={N^2 \o R^7 M_{pl}^{21}} {v^8\o r^{15}}=
 G^2_{10} {N^2 \o M_{pl}^3 R^5} 
{v^8\o r^{15}},}
where $M_{pl}$ is the eleven-dimensional Planck mass.
We would like to add this correction to the Lagrangian {\bi}
and compute the result for the gravitational interaction.
Using the same formulas that we used for the leading term 
and the relativistic corrections we obtain 
\eqn\bx{ V(r) \sim G^2_{10} {1 \o R^3 M_{pl}^3} {M^2 \o r^{15}}.}
We can rewrite this expression in terms of the string
coupling constant $g_s$ and the ten-dimensional
gravitational constant.
Recall that the eleven-dimensional Planck length 
is related to the string 
coupling constant and $\a'$ as

\eqn\bxi{l_{pl}=( 2\pi g_s)^{1/3} \sqrt{\a'},}
while the compactification radius is
\eqn\bxii{R=g_s \sqrt{\a'}.}
This means that the string coupling constant can be
expressed in terms of $R$ and $M_{pl}$ as
\eqn\bxiii{g^2_s={2 \pi R^3 \o l_{pl}^3}=2 \pi R^3 M_{pl}^3.}
From this it follows
\eqn\bxxxx{G_{10} \sim {G_{11} \o R} \sim g_s^2 {\a'}^4.}
The dimensionless coupling constant appearing in 
the correction to Newton's potential {\bx} is
nothing but the string coupling constant
\eqn\bxiv{V(r) \sim g_s^2{\a'}^8 {M^2 \o r^{15}} 
\sim {G_{10}^2 \o g_s^2} {M^2 \o r^{15}}.}
This result looks similar to the finite quantum gravity correction
appearing in the four-dimensional formula {\aiii}.
In string theory conventions we can set $\hbar =1$ or equivalently
$\hbar$ can be shifted away with a scale tranformation of the metric.
This is the origin of the dependence on the string coupling constant
in the above formula. We will see this in more detail in 
the next section.
The fact that integer powers of Newton's constant and $g_s$ appear
is indicating that this is a finite correction from the supergravity
point of view. Otherwise fractional powers of $G_{10}$ would be
present due to the existence of a dimensionful parameter, the cutoff.
The above quantum gravity correction is the leading quantum effect
as other quantum effects are higher order in $1/r$.

To summarize, the leading terms of the eleven-dimensional
Newton's potential from the Matrix 
theory point of view take the form
\eqn\bxv{V(r) \sim {G_{10} M^2 \o r^7} 
\left(1+{G_{10} M \o r^7} +{G_{10} \o g_s^2} {1 \o r^8} + \dots\right).}
This formula has an interesting analogy with {\aiii}.
The different powers in $r$ are due to the fact that 
the dimensions of Newton's constant are different in 
ten and in four dimensions.
Let us now have a closer look at the supergravity
side of the story.

\newsec{The Newtonian Potential from Supergravity}
On the supergravity side we can follow closely the four-dimensional
calculation {\do} though we will not 
determine any numerical coefficients. Our aim is to illustrate
which Feynman diagrams contribute to the different
terms in the gravitational potential.

In ten dimensions 
two massive objects with mass $M$ separated
by a distance $r$ interact to lowest order
by the Newtonian potential
\eqn\di{V(r) \sim {G_{10} M^2 \o r^7}.}
If the mass or velocities of these objects get too big
the potential receives relativistic corrections.
These would be of
the form

\eqn\dii{V(r) \sim {G_{10} M^2 \o r^7} \left( 1+a {G_{10}M \o r^7}+\dots\right).}
The numerical coefficient $a$ is calculable 
in a post-Newtonian expansion. At some point this expression will
be corrected by quantum gravity effects. It is possible to figure out
the general form of these corrections using dimensional analysis.
Since they arise from loop diagrams they will involve an
additional power of Newton's constant $G_{10}$ and if they are 
quantum corrections they will be at least linear in $\hbar$.
As we have already mentioned, we will be interested in a 
very particular quantum gravity correction:
the leading long distance quantum gravity effect.
It will be a non-analytic effect in the momentum transfer, since 
analytic effects correspond to contact terms in coordinate space.
Since this effect is due to long range propagation of massless
particles, the other dimensionful parameter
is the distance $r$. The ten-dimensional Newton's constant
has dimension $M^{-1} L^7$ and $\hbar$ 
has dimensions $M L$.
The combination 
\eqn\diii{ {G_{10} \hbar \o r^8},}
is dimensionless and provides an expansion parameter 
for the long distance quantum effects.
Altogether, to leading order 
we expect a modification of the classical potential of 
the form

\eqn\div{V(r) \sim {G_{10} M^2 \o r^7} \left( 1+a {G_{10} M \o r^7} +b {G_{10}
 \hbar \o r^8}+\dots \right).}
The free parameters $a$ and $b$ are finite constants that 
can be calculated in a quantum theory of gravity by
computing Feynman diagrams. Let us illustrate the general idea.

Our starting point is the ten-dimensional Einstein action\foot{We will
not be taking fermions nor any other fields of type IIA supergravity
besides the graviton into account. These fields will have to be
taken into account to determine the precise numerical coefficients
of {\div}.} 
\eqn\ci{S_{grav} \sim {1 \o G_{10}} \int d^{10} x \sqrt{-g} R, }
coupled to a massive scalar field
\eqn\cii{S_{matter} \sim \int  d^{10}x \sqrt{-g} \left( {1 \o 2} 
g^{\mu \nu} \partial_{\mu} \varphi
\partial_{\nu} \varphi -{1 \o 2} M^2 \varphi^2 \right) }
The quantum fluctuations of the gravitational field $h_{\m\n}$ can be
expanded around a flat background 
\ref\thv{G.~`t Hooft and M.~Veltman, ``One Loop Divergencies
in the Theory of Gravitation'', {\it Ann. Poincare Phys. Theor.}
{\bf A20} (1974) 69.}
\eqn\ciii{g_{\mu \nu} =\eta_{\mu \nu} +\k_{10} h_{\mu \nu},}
where $\k_{10}^2=G_{10}$ up to a numerical coefficient.
To quantize the fluctuations one needs to fix a gauge and a gauge fixing
term has to be added to the Lagrangian.
One can then follow the standard rules to quantize the theory by 
computing Feynman diagrams.
We are interested in considering quantum corrections
at one loop. For that purpose the Lagrangians have to be expanded 
to quartic order in the fields. 

The leading term of Newton's potential follows from the single
graviton  exchange diagram where two scalar particles interact
through the exchange of a single graviton
\eqn\civ{M_{12}\sim \kappa_{10}^2 T_{\mu \nu}^{(1)}(q) D^{\mu \nu \alpha\beta} 
T_{\alpha \beta}^{(2)} (-q).}
Here $T_{\m\n}$ is the on-shell matrix element 
of the matter energy-momentum tensor and $D^{\m\n\a\b}$ is
the graviton propagator
\ref\dwitt{B.~S.~De Witt, ``Quantum Gravity III: Applications
of the Covariant Theory'', \physrev {162} {1966} {162}.}.
The Newton potential can then be found by
Fourier transforming

\eqn\cv{{1 \o M^2} M_{12} \sim{G_{10} M^2 \o q^2},}
where the factor $1/M^2$ on the left hand side
accounts for the proper normalization.
In coordinate space we obtain
\eqn\cvi{V(r) \sim G_{10} \int {d^9 q \o (2 \pi)^9} 
e^{-i q r} {M^2 \o q^2} \sim {G_{10} M^2 \o r^7}.}

To compute one loop corrections to Newton's potential 
we have to consider both vertex corrections and vacuum
polarization effects at one loop {\do}.
Double graviton exchange diagrams contribute to this order as well
\ref\ham{H.~W.~Hamber and S.~Liu, ``On the Quantum Corrections
to the Newtonian Potential'', \pl{357} {1995} {51}, hep-th/9505182.}.
The one loop diagrams will have an additional power 
of $\k_{10}^2$ compared to the tree diagram and $\k_{10}^2$ has dimension
$M^{-8}$. The combination 
$\k_{10}^2 q^8$ is dimensionless. However loop diagrams
will also produce non-analytic terms 
of the form $log(-q^2)$ and $\sqrt{{M^2\o {-q^2}}}$
which are dimensionless\foot{The constant which
makes the argument of the logarithm dimensionless
gives a contact term in coordinate space which we ignore in our discussion.}.
Such non-analytic effects lead to a power law behavior 
in coordinate space
\eqn\cvii{\eqalign{&
\int {d^9 q \o (2 \pi )^9} e^{-i qr} (\sqrt{-q^2})^5  \sim
{1\o r^{14}},\cr
& \int {d^9 q \o (2 \pi)^9} e^{-iqr} ( \sqrt{-q^2})^6 \log (-q^2)
 \sim {1\o r^{15}}.\cr}
}
On the other hand analytic terms correspond to delta functions
in coordinate space
\eqn\cviii{\int {d^9 q \o (2 \pi)^9} e^{-iqr} =\delta^9(r).}

We are interested in precisely the terms {\cvii} appearing
as one-loop effects. The first term will correspond to the
relativistic correction while the second term corresponds to
the leading quantum correction.
In momentum space the contributions to the potential take
the form
\eqn\civ{V(q) \sim G_{10}M^2\left[ {1 \o q^2}+G_{10}
\left( (\sqrt{-q^2})^6\log(-q^2) +\sqrt{-q^2}^5M\right)\right], }
with some finite numerical coefficients that can be calculated.
After Fourier transforming we obtain the Newtonian potential

\eqn\dx{V(r) \sim {G_{10} M^2 \o r^7} \left( 1+a {G_{10} M \o r^7} +b { 
G_{10}\hbar
  \o r^8}+\dots \right).}
Here we have restored the $\hbar$ dependence by dimensional analysis.
This result takes the same form as {\bxv}. Of course in section 2 
we have been using conventions 
where ${\hbar}=1$ as one usually does in string theory.
We can absorb the value of ${\hbar}$ 
in ${\dx}$ with a rescaling of the metric
\ref\gsw{M.~B.~Green, J.~H.~Schwarz and E.~Witten, 
``Superstring Theory'',  Volume 2, page 326, 
Cambridge University Press, 1987.}. This will
introduce the dilaton dependence.
Consider the action {\ci}. 
Under a scale transformation where lengths are rescaled by
a factor $t^{-1}$ the metric transforms as
\eqn\ei{g_{\mu \nu}\rightarrow t^{-2} g_{\mu \nu}. }
Under this transformation the curvature $R_{\m\n}$ will remain
invariant so that the ten-dimensional action scales as
\eqn\eii{S\rightarrow t^{-8} S.}
At the classical level the normalization of the action is not
relevant so that general relativity is scale invariant at this level.
When we pass to the quantum theory we compute a path integral
\eqn\eiii{Z=\int e^{{i \o \hbar} S}.}
The parameter $\hbar$ appearing in this expression can
be absorbed into a transformation of the type {\ei}. In other words, 
under a scale tranformation ${\hbar}$ will transform as
${\hbar}\rightarrow t^{-8}{\hbar}$.
Furthermore, type IIA supergravity 
has the same classical scale invariance if the 
dilaton behaves in the following way under scale transformations
\eqn\eiv{ \phi \rightarrow t^2 \phi.}
From these considerations it follows that 
the effect of order ${\hbar}$ that we have been considering
corresponds to a $\phi^{-4}$-effect in type IIA supergravity,
or equivalently a $1/g_s^2$-effect. 
This is what we wanted to show.

\newsec{Conclusions}
In this paper we have shown that the Matrix model
for Schwarzschild black holes of {\hm} and {\bfks} 
describes correctly the leading long range quantum gravity correction
to Newton's potential in eleven dimensions. This effect is finite
and can be computed from the supergravity point of view from one-loop
Feynman diagrams. We have found agreement up to numerical factors that
we have not calculated. 
We have further shown that relativistic corrections to Newton's potential
are correctly reproduced.
Obviously it would be of interest to know if it is possible to match 
the precise numerical coefficients between the Matrix model and supergravity.
Since the model of {\hm} and {\bfks} purports to describe 
Schwarzschild black holes in diverse dimensions as toroidal compactifications
of the eleven-dimensional model it would be interesting to know
if the lower dimensional models also correctly reproduce the finite
quantum gravity correction to Newton's potential.

\vskip 1cm
\noindent {\bf Acknowledgement}

\noindent 
We would like to thank J.~Schwarz for discussions 
and I.~Klebanov for correspondence.
This work was supported by the U.S. Department of Energy 
under grant DE-FG03-92-ER40701.

\listrefs

\end